\documentclass[aps,prb,twocolumn,showpacs]{revtex4-1}

\usepackage{graphicx}
\usepackage{amsmath}
\begin{document}

\title{Combined first-principles and EXAFS study of structural instability
in BaZrO$_3$}

\author{A. I. Lebedev}
\email[]{swan@scon155.phys.msu.ru}
\author{I. A. Sluchinskaya}
\affiliation{Physics Department, Moscow State University, Moscow, 119991, Russia}

\date{\today}

\begin{abstract}
Phonon spectrum of cubic barium zirconate is calculated from first principles
using the density functional theory. Unstable phonon mode with the $R_{25}$
symmetry in the phonon spectrum indicates an instability of the cubic structure
with respect to rotations of the oxygen octahedra. It is shown that the
ground-state structure of the crystal is $I4/mcm$. In order to find the
manifestations of the predicted instability, EXAFS measurements at the Ba
$L_{\rm III}$-edge are used to study the local structure of BaZrO$_3$ at 300~K.
An enhanced value of the Debye--Waller factor for the Ba--O atomic pair
($\sigma^2_1 \sim 0.015$~{\AA}$^2$) revealed in the experiment is associated
with the predicted structural instability. The average amplitude of the thermal
octahedra rotation estimated from the measured $\sigma^2_1$ value is
$\sim$4~degrees at 300~K. The closeness of the calculated energies of different
distorted phases resulting from the condensation of the $R_{25}$ mode suggests
a possible structural glass formation in BaZrO$_3$ when lowering temperature,
which explains the cause of the discrepancy between the calculations and
experiment.

\end{abstract}


\maketitle

\section{\label{sec1}Introduction}

Crystals with a general formula $AB$O$_3$ and the perovskite structure
form a large family whose characteristic feature is a tendency to various
(ferroelectric, ferroelastic, magnetic, superconducting) phase transitions.
Barium zirconate BaZrO$_3$, one of the members of this family, is a crystal
with a high melting point ($\sim$2600$^\circ$C), moderate dielectric constant
($\sim$40), low dielectric losses, and low thermal expansion coefficient.
It is widely used as a material for microwave ceramic capacitors and for
inert crucibles used for growth of single crystals of high-temperature
superconductors. In recent years, the interest to barium zirconate has revived
again because it was found that BaZrO$_3$ doped with acceptor impurities, in
particular yttrium, is a high-temperature ionic conductor with high proton
conductivity, which can be used in solid-oxide fuel
cells.~\cite{AnnuRevMaterRes.33.333,JAmCeramSoc.88.2362,JMaterChem.20.6265,
NatureMater.9.846}  Its solid solutions with BaTiO$_3$ are relaxors with a
very high electric-field tuning of the dielectric constant;~\cite{ActaMater.52.5177,
ApplPhysLett.90.182901,JAdvDielectrics.2.1230002}  this property can be used
in tunable filters, oscillators, phase shifters, etc. It should be noted that
up to now, the properties of BaZrO$_3$ have been studied mainly at room and
higher temperatures. Only a few experiments~\cite{PhysRevB.72.205104} (in
addition to the first-principles calculations~\cite{PhysRevLett.74.2587,
PhysRevB.72.205104,PhysRevB.73.180102,PhysRevB.78.012103,PhysRevB.79.174107,
JApplPhys.105.044110}) were undertaken to study the properties of BaZrO$_3$
(the ground-state structure, the behavior of the dielectric constant, etc.)
at low temperatures at which most perovskite oxides exhibit interesting
properties.

Recent X-ray and neutron diffraction studies~\cite{PhysRevB.72.205104} did not
found any phase transitions in BaZrO$_3$ at $T < 300$~K, so that barium zirconate
retains the cubic $Pm3m$ structure down to very low temperatures (2~K). These
structural data, however, disagree with the results of the first-principles
phonon spectra calculations for BaZrO$_3$ (Refs.~\onlinecite{PhysRevLett.74.2587,
PhysRevB.72.205104,PhysRevB.73.180102,PhysRevB.79.174107,JApplPhys.105.044110}),
in which the appearance of the unstable $R_{25}$ mode indicates an instability
of the cubic structure with respect to the rotation of the oxygen octahedra.
In Ref.~\onlinecite{PhysRevB.73.180102}, the equilibrium low-symmetry $P\bar{1}$
structure was found, which results from the doubling the unit cell in all
directions and the rotation of the ZrO$_6$ octahedra according to Glazer tilt
system $a^-b^-c^-$. Taking into account of these rotations was shown to improve
the agreement between the calculated values of the low-temperature dielectric
constant with experiment.~\cite{PhysRevB.73.180102} An indirect indication of
the possible phase transformation in BaZrO$_3$ crystals can be its very low
thermal expansion coefficient at $T < 300$~K.~\cite{PhysRevB.72.205104}
In order to explain the disagreement between the theory and experiment, it was
suggested that, for some reason, the long-range order in the octahedra rotations
in these crystals cannot be established even at low temperatures. In particular,
it was supposed that it may be caused by zero-point lattice
vibrations.~\cite{PhysRevLett.74.2587,PhysRevB.72.205104,PhysRevB.78.012103}

The aim of this work is to calculate the phonon spectrum and the ground-state
structure for BaZrO$_3$ from first principles and to find experimental evidence
for the structural instability in this material. We started from the assumption
that, even if the long-range order in barium zirconate is absent for some reason,
the structural instability, if it exists, should be manifested in the change of
the \emph{local structure}. To search for these changes, the extended X-ray
absorption fine structure (EXAFS) studies are used. EXAFS is one of the modern
powerful experimental techniques that enables to obtain detailed information on
the local structure of crystals.~\cite{RevModPhys.53.769}  We expected to find
the evidence for the structural instability in BaZrO$_3$ from anomalous behavior
of the Debye--Waller factors which characterize the magnitude of the local
displacements and the amplitude of thermal vibrations.

\section{Calculation and experimental techniques}

In this work, the calculations were performed from first principles using the
density functional theory. The pseudopotentials for Ba, Zr, and O atoms used
in the calculation were borrowed from Refs.~\onlinecite{PhysSolidState.51.362,
PhysSolidState.52.1448}.  The cut-off energy
of plane waves was 30~Ha (816~eV). The integration over the Brillouin zone was
performed on the 8$\times$8$\times$8 Monkhorst--Pack mesh. The lattice
parameters and the equilibrium atomic positions were determined from the
condition of decreasing of the residual forces acting on the atoms below
$5 \cdot 10^{-6}$~Ha/Bohr
(0.25~meV/{\AA}) while the accuracy of the total energy calculation was better
than 10$^{-10}$~Ha. To calculate of the phonon spectra, the formulas derived
from the density functional perturbation theory and the interpolation
technique~\cite{PhysSolidState.51.362} were used. The calculated
lattice parameter for the cubic BaZrO$_3$ was $a_0 = 4.1659$~{\AA} and
agreed well with the experimental value obtained at 10~K (4.191~{\AA},
Ref.~\onlinecite{PhysRevB.72.205104}). The slight difference of these values
was due to a systematic underestimation of the lattice parameter typical for
the local density approximation (LDA) used in this work.

The sample of barium zirconate was prepared by the solid phase reaction method.
The initial components were BaCO$_3$ and microcrystalline ZrO$_2$ obtained by the
decomposition of ZrOCl$_2$\,$\cdot$\,8H$_2$O at 300$^\circ$C. The components were
dried at 600$^\circ$C, weighed in proper proportions, ground in
acetone and calcined in air at 1100$^\circ$C for 6~h. The resulting powder was
ground again and annealed at 1500$^\circ$C for 3~h. The single-phase character
of the sample was confirmed by X-ray diffraction.

Four EXAFS spectra were recorded using the simultaneous registration of the
transmission signal and X-ray fluorescence on the KMC-2 station of the BESSY
synchrotron radiation source (the beam energy of 1.7~GeV, the beam current up
to 300~mA) at the Ba $L_{\rm III}$-edge (5.247~keV) at 300~K. The intensity of
the incident radiation was measured by an ionization chamber, the intensity of
radiation transmitted through the sample was measured with a silicon photodiode,
and the intensity of the X-ray fluorescence excited in the sample was measured
using the R\"ONTEC silicon energy-dispersive detector.

The Ba $L_{\rm III}$-edge was chosen for measurements because the expected
structural instability, which is accompanied by the rotation of the oxygen
octahedra, should manifest most strongly in the change of the Ba--O
interatomic distances, and not in the Zr--O interatomic distances. This is
because the Zr--O distances change very little when the octahedra are rotated.
In principle, the rotation of the octahedra could be also observed at the Zr
$K$-edge as the changes in the multiple-scattering contributions to the EXAFS
spectra as proposed in Ref.~\onlinecite{PhysRevLett.72.1352}.

The EXAFS spectra processing was carried out in the traditional
way.~\cite{PhysRevB.55.14770}  The spectra were independently proceeded, and
the results were then averaged. Since, according to our theoretical predictions,
the temperature of the expected phase transition is below 300~K, the structure
of BaZrO$_3$ at room temperature is cubic, and so the structural instability
should be manifested as enhanced Debye--Waller factors.

\section{Results of calculations}

\begin{figure}
\centering
\includegraphics{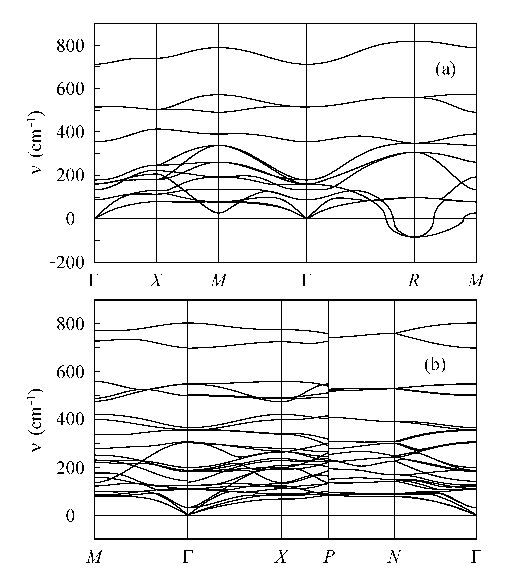}
\caption{Phonon spectrum of BaZrO$_3$ in the cubic $Pm3m$ phase (\emph{a}) and
the ground-state $I4/mcm$ phase (\emph{b}).}
\label{fig1}
\end{figure}

The calculated phonon spectrum of BaZrO$_3$ with the cubic perovskite structure
(space group $Pm3m$) is shown in Fig.~\ref{fig1}a. It is seen that an unstable
phonon with $R_{25}$ symmetry (imaginary phonon energies are represented in the
figure as negative values) appears in this spectrum. The eigenvector of the
unstable phonon indicates that
the cubic structure is unstable with respect to the octahedra rotations. The
results of our calculations are consistent with the results of previous
calculations,~\cite{PhysRevLett.74.2587,PhysRevB.72.205104,PhysRevB.73.180102,
PhysRevB.79.174107,JApplPhys.105.044110}  but it should be noted that an
additional instability at the $M$ point, which was observed
in Ref.~\onlinecite{PhysRevB.79.174107}, was not found in our phonon spectrum.

\begin{table}
\caption{\label{table1}The energies of different distorted phases of
BaZrO$_3$. The energy of the cubic $Pm3m$ phase was taken as the energy
reference.}
\begin{ruledtabular}
\begin{tabular}{cccc}
Unstable   &  Glazer    & Space group    & Energy \\
mode       &  rotations &                & (meV)  \\
\hline
\rule{0pt}{4mm}%
$R_{25}$ & $a^-a^-a^-$  & $R{\bar 3}c$  & $-$9.17 \\
$R_{25}$ & $a^0b^-b^-$  & $Imma$        & $-$9.47 \\
$R_{25}$ & $a^0a^0c^-$  & $I4/mcm$      & $-$10.01 \\
\end{tabular}
\end{ruledtabular}
\end{table}

To determine the structure of the ground state, the energies of different
distorted phases resulting from the condensation of triply degenerate unstable
$R_{25}$ mode were calculated. We write a Landau-type expansion of the total
energy of a crystal in a power series of the amplitudes of the distortions,
\begin{equation}
\begin{split}
E_{\rm tot} = E_{\rm tot}(0) + a(\phi_x^2 + \phi_y^2 + \phi_z^2) + \\
b(\phi_x^2 + \phi_y^2 + \phi_z^2)^2 + c(\phi_x^4 + \phi_y^4 + \phi_z^4) + \\
d(\phi_x^2 + \phi_y^2 + \phi_z^2)^3 + e \phi_x^2 \phi_y^2 \phi_z^2 + ...~,
\end{split}
\end{equation}
where $E_{\rm tot}(0)$ is the total energy of the parent cubic phase, and
$\phi_x$, $\phi_y$, and $\phi_z$ are the angles of rotation around three
fourfold axes of the cubic structure. From this expansion it can be shown that
to find the ground-state structure, it is sufficient to calculate the energies
of phases described by the order parameters ($\phi$,0,0), ($\phi$,$\phi$,0), and
($\phi$,$\phi$,$\phi$), which correspond to the minimums of $E_{\rm tot}$ for
different combinations of signs of the coefficients $c$ and $e$. The specified
order parameters correspond to the Glazer tilt systems $a^0a^0c^-$, $a^0b^-b^-$,
and $a^-a^-a^-$, and lead to the $I4/mcm$, $Imma$, and $R{\bar 3}c$ space groups,
respectively. As follows from Table~\ref{table1}, the $I4/mcm$ phase has the
lowest energy among these phases. This phase can be obtained from the cubic
$Pm3m$ structure by the out-of-phase rotation of the oxygen octahedra around one
of the fourfold axes. The energy of this phase is $\sim$10~meV lower than the
energy of the cubic phase; this gives an estimated phase transition temperature
of $\sim$120~K. The calculations of the phonon spectrum (Fig.~\ref{fig1}b) and
elastic properties for the $I4/mcm$ phase prove that this phase is the
ground-state structure.

It should be noted that among the obtained phases there is no $P{\bar 1}$
solution, which was considered as the ground-state structure in
Ref.~\onlinecite{PhysRevB.73.180102}.
This phase corresponds to the Glazer tilt system $a^-b^-c^-$, in which the
rotation angles around three fourfold axes of the cubic structure are different.
The testing of this solution showed that the $P{\bar 1}$ structure very slowly
relaxes to the $I4/mcm$ solution found above (the convergence to this
result required over 400~iterations). This demonstrates
that the use of the information on the symmetry of the total energy
can significantly (by more than 10~times) reduce the computation time
needed to find the ground-state structure and guarantees a correct result.

\begin{table*}
\caption{\label{table2}Calculated lattice parameters and atomic coordinates
in the ground-state structure of BaZrO$_3$.}
\begin{ruledtabular}
\begin{tabular}{ccccccc}
Phase   & Lattice       & Atom & Wyckoff  & $x$     & $y$     & $z$ \\
        & parameters ({\AA}) & & position &         &         & \\
\hline
$I4/mcm$ & $a = 5.8722$ & Ba   & $4b$    & 0.00000 & 0.50000 & 0.25000 \\
         & $c = 8.3577$ & Zr   & $4c$    & 0.00000 & 0.00000 & 0.00000 \\
         &              & O    & $4a$    & 0.00000 & 0.00000 & 0.25000 \\
         &              & O    & $8h$    & 0.22315 & 0.72315 & 0.00000 \\
\end{tabular}
\end{ruledtabular}
\end{table*}

The lattice parameters and atomic coordinates in the ground-state structure
of BaZrO$_3$ are given in Table~\ref{table2}.

\section{Experimental results}

\begin{table*}
\caption{\label{table3}Structural parameters for the first three shells
in cubic BaZrO$_3$ at 300~K obtained from the EXAFS data.}
\begin{ruledtabular}
\begin{tabular}{ccccccc}
Measurement method & $R_1$~({\AA}) & $\sigma_1^2$~({\AA}$^2$) & $R_2$~({\AA}) & $\sigma_2^2$~({\AA}$^2$) & $R_3$~({\AA}) & $\sigma_3^2$~({\AA}$^2$) \\
\hline
\rule{0pt}{4mm}%
Fluorescence  & 2.916 & 0.0155 & 3.640 & 0.0079 & 4.244 & 0.0101 \\
Transmission  & 2.914 & 0.0145 & 3.635 & 0.0068 & 4.244 & 0.0087 \\
\end{tabular}
\end{ruledtabular}
\end{table*}

\begin{figure}
\centering
\includegraphics{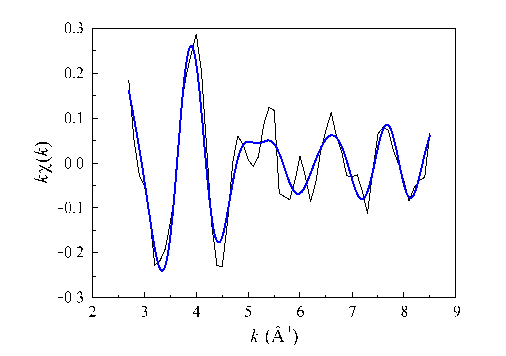}
\caption{(Color online) A typical EXAFS spectrum of BaZrO$_3$ obtained at the
Ba $L_{\rm III}$-edge at 300~K (thin, black line) and the results of its fitting
(thick, blue line).}
\label{fig2}
\end{figure}

A typical EXAFS spectrum for the BaZrO$_3$ sample obtained at the Ba
$L_{\rm III}$-edge at 300~K is shown in Fig.~\ref{fig2}. The data analysis
shows that the spectra are consistent with a model in which the structure of
the sample is cubic. The structural parameters (interatomic distances and
the Debye--Waller factors) for the first three shells are given in
Table~\ref{table3}. It is seen that the structural parameters obtained using two
different measurement methods (transmission and fluorescence) are practically
the same.

In Table~\ref{table3}, it should be noted that the Debye--Waller factor for
the first shell is much higher than the values for the second and third shells
(usually the values of this factor increase monotonically with increasing
interatomic distance).

The values of the Debye--Waller factors are determined by two components:
(1)~static distortions of the structure and (2)~thermal vibrations. As our
measurements are performed at a temperature above the temperature of the
expected phase transition, the static distortions of the structure are absent.
Furthermore, the optical modes responsible for the deformation of the oxygen
octahedron are high-energy modes in the phonon spectrum of BaZrO$_3$ and are not
excited at 300~K. Therefore, we can consider the octahedron ZrO$_6$ as a rigid
unit. The thermal motion of the center of the octahedron relative to the Ba atom
is characterized by the Debye--Waller factor $\sigma^2_2$. Since (as result of
the rigidity of the octahedron) the oxygen atoms are strongly bound to the Zr
atoms, the only way to explain why $\sigma^2_1 \gg \sigma^2_2$ is to take into
account the rotation of the octahedra. Assuming that in the absence of the
octahedra rotation the Debye--Waller factors for the Ba--O and Ba--Zr bonds are
close (0.008~{\AA}$^2$), the excess part of the Debye--Waller factor in the
first shell, $\Delta \sigma^2_1 = \sigma^2_1 - \sigma^2_2$, can be associated
with the thermal rotations. Then, using the formula
$$ \Delta \sigma^2_1 = \frac{a^2_0 \theta^2}{12}\,, $$
which describes the dependence of $\Delta \sigma^2$ on the octahedra rotation
angle $\theta$ about the fourfold axis of the cubic structure, we can estimate
the amplitude of the thermal rotations. For the lattice parameter of
$a_0 \approx 4.2$~{\AA}, the mean rotation angle is
$\sqrt{{\bar {\theta}}^2} \approx 4^\circ$. Taking into account the results of
Ref.~\onlinecite{PhysRevB.60.2961} in which the experimental value of the
rotation angle in SrTiO$_3$ at 50~K (well below the temperature of the
structural phase transition) is 2.01$^\circ$, we can conclude the existence of
the structural instability in BaZrO$_3$.

\section{Discussion}

Now, we discuss why the long-range order in the octahedra rotations is absent
in BaZrO$_3$ at low temperatures. As was mentioned in Sec.~\ref{sec1}, one of
the reasons for this behavior was the suppression of the long-range order by
zero-point vibrations. In Ref.~\onlinecite{PhysSolidState.51.802}, a criterion
was proposed, which enables, using a set of parameters obtained entirely from
first principles, to estimate the stability of a structure with respect to
zero-point vibrations. According to this criterion, zero-point vibrations
prevent the system to be localized in one of the local minima of the potential
if $h \nu / E_0 > 2.419$, where $\nu$ is the unstable phonon frequency and $E_0$
is the depth of the local minimum of the potential. In our calculations, for
the unstable $R_{25}$ mode $h \nu = 81.2i~{\rm cm}^{-1} \approx 10$~meV and
$E_0 \approx 10$~meV (Table~\ref{table1}); therefore, zero-point vibrations
should not suppress the structural distortions in barium zirconate. The phonon
frequency of 79$i$~cm$^{-1}$ and the local minimum depth of 8~meV obtained for
BaZrO$_3$ in Ref.~\onlinecite{PhysRevB.73.180102}  give the ratio
$h \nu / E_0 \approx 1.2$, which is close to our value. The parameters
$h \nu \approx 27i$~cm$^{-1}$ and $E_0 = 1.5$~meV obtained in
Ref.~\onlinecite{PhysRevB.72.205104}  are much lower and give
$h \nu / E_0 \approx 2.2$, which is anyway lower then the critical value.
Therefore, the explanation proposed in
Ref.~\onlinecite{PhysRevLett.74.2587,PhysRevB.72.205104,PhysRevB.78.012103} that
the suppression of the long-range order in BaZrO$_3$ is caused by zero-point
vibrations seems unlikely. The fact that the influence of zero-point vibrations
on the phase transition in BaZrO$_3$ is weaker than in SrTiO$_3$ is not
surprising because the masses of both metal atoms in barium zirconate are
larger.

\begin{table}
\caption{\label{table4}The minimum and maximum energies of the phases resulting
from the condensation of rotational $R_{25}$ and $M_3$ modes in some perovskite
crystals and the difference of these energies.}
\begin{ruledtabular}
\begin{tabular}{ccccc}
Compound & Unstable & Space & Energy  & $\Delta E$ \\
         & mode     & group & (meV)   & (meV) \\
\hline
BaZrO$_3$ & $R_{25}$ & $R{\bar 3}c$ & $-$9.17 & 0.84 \\
          & $R_{25}$ & $I4/mcm$     & $-$10.01 \\
SrTiO$_3$ & $M_3$    & $P4/mbm$     & $-$9.5  & 21.4~\cite{PhysSolidState.51.362} \\
          & $R_{25}$ & $I4/mcm$     & $-$30.9 \\
CaTiO$_3$ & $M_3$    & $P4/mbm$     & $-$321  & 176~\cite{PhysSolidState.51.362} \\
          & $R_{25} + M_3$ & $Pbnm$ & $-$497 \\
CdTiO$_3$ & $R_{25}$ & $I4/mcm$     & $-$912  & 371~\cite{PhysSolidState.51.802} \\
          & $R_{25} + M_3$ & $Pbnm$ & $-$1283 \\
SrZrO$_3$ & $R_{25}$ & $I4/mcm$     & $-$298  & 69 \\
          & $R_{25} + M_3$ & $Pbnm$ & $-$367 \\
\end{tabular}
\end{ruledtabular}
\end{table}

The theoretical calculations performed in this work enable to propose another
explanation for the absence of the long-range order in BaZrO$_3$ crystals in
which the structural instability is expected. The closeness of the total
energies of the $R{\bar 3}c$, $Imma$, and $I4/mcm$ phases (Table~\ref{table1}),
which can result from the condensation of the $R_{25}$ mode, but do not satisfy
the group--subgroup relation, enables to suggest the appearance of the
structural glass state when cooling the crystals. The internal strains and
defects cause the appearance of different phases with close energies at
different points of the sample, obviously resulting in the absence of the
long-range order in the octahedra rotations. Moreover, the oxygen vacancies,
which can be generated in BaZrO$_3$ quite easily, disturb the three-dimensional
connectivity of the perovskite structure and can also contribute to the
structural glass formation. The uniqueness of barium zirconate among other
crystals with the perovskite structure results from the fact that the energy
difference between the phases with different rotation patterns is only 0.84~meV
(Table~\ref{table1}), which is much less than in other investigated crystals
(Table~\ref{table4}). In our opinion, this small energy difference may explain
the tendency of BaZrO$_3$ to the formation of the structural glass.

Further experiments on BaZrO$_3$, such as studies of the temperature dependence
of the diffraction line width, diffuse scattering, elastic properties, and
other physical phenomena sensitive to the rotations of the octahedra, could
provide additional information on the cause of the discrepancy between theory
and experiment and to confirm or refute our explanation. We think that the
low-temperature Raman studies of barium zirconate can also be very useful; our
calculations show that five first-order Raman lines located at 16--36, 104--111,
121--128, 350--361, and 547--548~cm$^{-1}$ appear in three possible distorted
phases ($R{\bar 3}c$, $Imma$, and $I4/mcm$) with the octahedra rotations.

\section{Conclusion}

First-principles calculations of the phonon spectrum of BaZrO$_3$ performed
in this work and experimental observation of the enhanced Debye--Waller factor
for the Ba--O bonds in the EXAFS study of BaZrO$_3$ suggest that barium
zirconate actually exhibits the structural instability associated with the
rotations of the oxygen octahedra. It is shown that zero-point lattice
vibrations are not the factor that prevents the establishing of the long-range
order in the octahedra rotations. Very close energies of the phases with the
different rotation patterns, which can result from the condensation of the
$R_{25}$ phonon, suggest a possible transition to the structural glass state
when cooling BaZrO$_3$ samples. This explains the apparent contradiction between
the theoretical predictions and the absence of the long-range order in the
octahedra rotations observed in the experiment.

\begin{acknowledgments}
The calculations presented in this work were performed on the laboratory
computer cluster (16~cores). The authors are grateful to the BESSY staff
and to the Russian-German laboratory for the financial support during their
stay at BESSY. This work was partially supported by the Russian Foundation
for Basic Research grant No. 13-02-00724.
\end{acknowledgments}


\providecommand{\BIBYu}{Yu}

\end{document}